# Generalized expressions for hyper-Rayleigh scattering from isotropic liquids

Yixing Chen[1], Sylvie Roke[1,#]

[1]Laboratory for fundamental BioPhotonics (LBP), Institute of Bio-engineering (IBI), and Institute of Materials Science (IMX), School of Engineering (STI), and Lausanne Centre for Ultrafast Science (LACUS), École Polytechnique Fédérale de Lausanne (EPFL), CH-1015, Lausanne, Switzerland, [#]sylvie.roke@epfl.ch;

**Abstract**

Generalized expressions for second harmonic scattering from isotropically distributed liquid molecules are derived for arbitrary scattering angles and polarization states.

**Introduction**

Second harmonic generation (SHG) is widely used as a powerful tool to probe the molecular structure and nonlinear optical properties of liquids, nanostructures and interfaces.[1-3] As a second-order nonlinear optical process, SHG is especially sensitive to both structural and electronic anisotropy in a material. A recent study of electrolyte solutions, using femtosecond elastic second harmonic scattering (SHS), has found that electrolytes induce long-range orientational order in water, starting at electrolyte concentrations as low as 10 μM, and the observed isotope effects indicate the important role of water hydrogen-bonding in the bulk water structuring.[4] These findings indicate that SHS from liquids consists of an incoherent and a coherent component. The coherent component originates only from correlated molecules. The second harmonic (SH) intensity is thus given by the following sum[5]

$$I(2\omega) \propto N \left\{ \langle \left(\beta_{ijk}^{(2)}\right)^2 \rangle_{incoh} + \langle \left(\beta_{ijk}^{(2)}\right)^2 \rangle_{coh} \right\} \quad (1)$$

in which $N$ is the number density of molecules, $\beta_{ijk}^{(2)}$ is the tensor element of the second-order hyperpolarizability $\boldsymbol{\beta}^{(2)}$. ⟨ ⟩ represents an ensemble orientational average over all involved molecules. *incoh* and *coh* denote the incoherent and coherent contribution to SHS respectively.

The results of Ref. [4] have initiated theoretical studies to understand the nature of the coherent term[6,7]. Although previous studies[8] have noted a deviation between SHS from liquids and the response of isotropically distributed molecules with a static geometry, the liquid has been considered as an isotropic distribution of molecules. The incoherent SHS from isotropic liquids is commonly referred to as hyper-Rayleigh scattering (HRS). Terhune *et al.*[9] reported on HRS from water in 1965 and in the same year Cyvin *et al.*[10] proposed the first basic theory of HRS. Following that, the theory of HRS was further developed in great detail by Bersohn *et al.*[5], Kauranen and Persoons[11], and other authors[12,13]: The work of Cyvin *et al.*[10] focused on the symmetry properties of $\boldsymbol{\beta}^{(2)}$ and its link to the HRS response of liquids. The geometry of the input and measured optical field was not considered. Bersohn *et al.*[5] developed a more general theory for HRS including considerations of the scattering geometry. Kauranen and Persoons[11] focused on using specific scattering geometries with non-collinear input of optical fields to retrieve values of the molecular hyperpolarizability from experiments. The work of Shelton[13] discussed HRS from transverse and longitudinal nonlocal modes and from local modes. Only two linear polarizations of the input light were considered.

In this work, we first present the derivation of general expressions of HRS intensities measured at arbitrary positions in 3-dimensional space with collinear incident laser beams of arbitrary polarizations. Then, we show general and complete expressions of HRS intensities from isotropic liquids in SI units with a discussion about independent non-zero $\boldsymbol{\beta}^{(2)}$ tensor elements. Differences with the abovementioned works are discussed and we provide a list of where we think errors or typos have appeared.

**Basic theory of second harmonic scattering from molecules**

We consider an experimental scheme as shown in Fig. 1. A laser beam with the frequency $\omega$ and wavevector $\boldsymbol{k}_1$ is focused into a liquid sample. We define, in the frequency domain, the electric field of the incident laser beam felt by molecule $v$ at position $\boldsymbol{r}$ as

$$\widetilde{\boldsymbol{E}}_v(\omega) \equiv \boldsymbol{E}(\omega)e^{i\boldsymbol{k}_1\cdot\boldsymbol{r}} = (E_x\hat{\boldsymbol{x}} + e^{i\phi_0}E_y\hat{\boldsymbol{y}})e^{i\boldsymbol{k}_1\cdot\boldsymbol{r}}, \qquad (2)$$

where $\hat{\boldsymbol{x}}$ and $\hat{\boldsymbol{y}}$ are the unit vectors along the x- and y-axis of the lab frame, respectively. $\phi_0$ is the phase difference between the two orthogonal components, $E_x$ and $E_y$, of the electric field and thus reflects the polarization state of the incident laser beam. Attenuation of the incident laser beam during propagation is neglected here. Linear polarization P (S) is defined as the linear polarization state of light field in the direction parallel (perpendicular) to the $xz$-scattering plane. Thus, $E_x \neq 0, E_y = 0$ for P polarization and $E_x = 0, E_y \neq 0$ for S polarization. Note that $(\omega)$ is omitted for $E_x$ and $E_y$ to simplify the notation.

For each molecule, the incident laser beam induces a second-order molecular dipole $\widetilde{\boldsymbol{p}}^{(2)}$ that oscillates at $2\omega$ and emits second harmonic light. For molecule $v$, the induced molecular dipole is given by[5]

$$\widetilde{\boldsymbol{p}}_v^{(2)} \equiv \boldsymbol{p}_v^{(2)}e^{i2\boldsymbol{k}_1\cdot\boldsymbol{r}} = \boldsymbol{\beta}_v^{(2)}:\widetilde{\boldsymbol{E}}_v(\omega)\widetilde{\boldsymbol{E}}_v(\omega) \qquad (3)$$

in which $\boldsymbol{\beta}_v^{(2)}$ represents the second-order hyperpolarizability of molecule $v$, a rank-3 tensor that characterizes the SH response of the molecule. Here $\boldsymbol{\beta}_v^{(2)}$ is given in the same coordinate system as $\widetilde{\boldsymbol{E}}_v(\omega)$, that is, the lab frame. Inserting Eq. 2 into Eq. 3, we obtain the three orthogonal components of the induced second-order molecular dipole as

$$p_{v,x}^{(2)} = \beta_{v,xyy}^{(2)}E_y^2 e^{i2\phi_0} + \beta_{v,xxx}^{(2)}E_x^2 + \left(\beta_{v,xyx}^{(2)} + \beta_{v,xxy}^{(2)}\right)E_xE_y e^{i\phi_0}$$

$$p_{v,y}^{(2)} = \beta_{v,yyy}^{(2)}E_y^2 e^{i2\phi_0} + \beta_{v,yxx}^{(2)}E_x^2 + \left(\beta_{v,yyx}^{(2)} + \beta_{v,yxy}^{(2)}\right)E_xE_y e^{i\phi_0}$$

$$p_{v,z}^{(2)} = \beta_{v,zyy}^{(2)}E_y^2 e^{i2\phi_0} + \beta_{v,zxx}^{(2)}E_x^2 + \left(\beta_{v,zyx}^{(2)} + \beta_{v,zxy}^{(2)}\right)E_xE_y e^{i\phi_0} \qquad (4)$$

where the subscript in $\boldsymbol{\beta}_v^{(2)}$ (e.g. $xyy$ in $\beta_{v,xyy}^{(2)}$) denotes the polarization state of $\widetilde{\boldsymbol{p}}_v^{(2)}$ and the two incident light field, respectively.

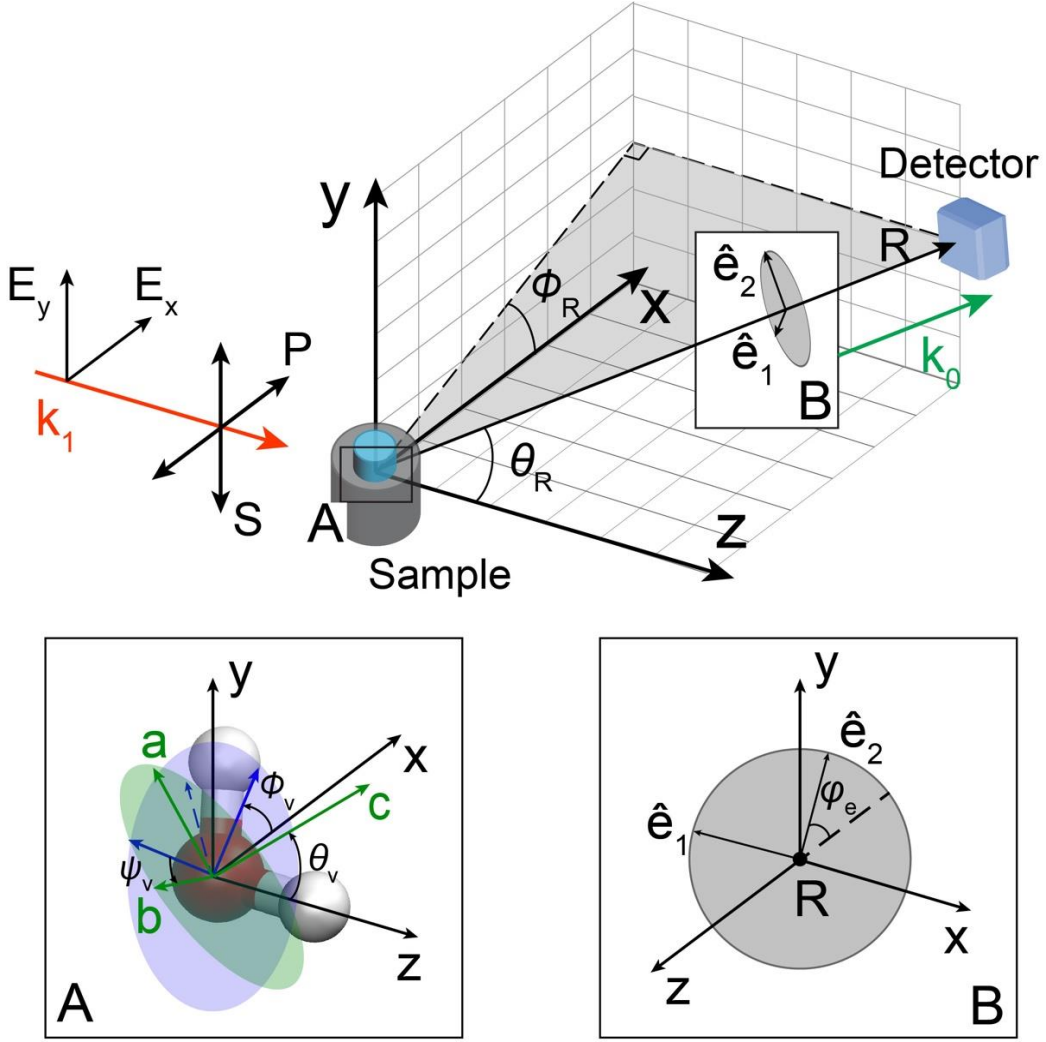

**Figure 1**. Schematic illustration of an HRS experiment. The Cartesian coordinate system $(x, y\, z)$ defines the lab frame and $(a, b, c)$ defines the molecular frame. The inset A illustrates the molecular frame $(a, b, c)$ of an individual water molecule. The lab frame and molecular frame are related by the Euler angles $\psi_v, \theta_v$, and $\phi_v$. The inset B shows a side-view of the polarization states of the emitted electric field, in the direction along the vector $-\mathbf{R}$ (which points into the paper). $\varphi_e$ is the angle between the unit vector $\hat{e}_2$ and the plane defined by $z$-axis and $\mathbf{R}$.

At the far-field observation position $\mathbf{R}$, where $|\mathbf{R}| \gg |\mathbf{r}|$, the electromagnetic field of the emitted SH light from the induced molecular dipole at position $\mathbf{r}'$ is given by[14]

$$\widetilde{\mathbf{E}}_v(2\omega) = \frac{k_{0,v}^2 e^{i\mathbf{k}_{0,v}\cdot(\mathbf{R}-\mathbf{r}')}}{4\pi\epsilon_0 R}\left(\widehat{\mathbf{R}} \times \widetilde{\mathbf{p}}_v^{(2)}\right) \times \widehat{\mathbf{R}} \tag{5}$$

where $k_{0,v}$ represents the magnitude of $\mathbf{k}_{0,v}$, the wavevector of the emitted SH light from the induced molecular dipole $\widetilde{\mathbf{p}}_v^{(2)}$. The direction of $\mathbf{k}_{0,v}$ is given by the unit vector $\widehat{\mathbf{k}}_{0,v} = \frac{\mathbf{R}-\mathbf{r}}{|\mathbf{R}-\mathbf{r}|} \sim \widehat{\mathbf{R}}$ for $|\mathbf{R}| \gg |\mathbf{r}|$. Therefore, for SHS in a small volume at the focus of the incident laser beam, the difference in the wavevector $\mathbf{k}_{0,v}$ of different molecules is negligible and we can approximate $\mathbf{k}_{0,v}$ as $\mathbf{k}_0$ for all molecules. $\widehat{\mathbf{R}}$ is the unit vector in the direction of $\mathbf{R}$ and is given by

$$\widehat{R} = \sin\theta_R \cos\phi_R \, \widehat{x} + \sin\theta_R \sin\phi_R \, \widehat{y} + \cos\theta_R \, \widehat{z} \qquad (6)$$

$\theta_R$ is the polar angle between $\widehat{R}$ and the $z$-axis and $\phi_R$ is the azimuthal angle between the $x$-axis and the projection of $\widehat{R}$ in the $xy$-plane as illustrated in Fig. 1. Inserting Eq. 6 into Eq. 5, we can expand the expression of $\widetilde{E}_v(2\omega)$ as

$\widetilde{E}_v(2\omega) \equiv E_v(2\omega) e^{i k_0 \cdot (R-r) + i 2 k_1 \cdot r}$, where $E_v(2\omega) \equiv E_{v,x}(2\omega)\widehat{x} + E_{v,y}(2\omega)\widehat{y} + E_{v,z}(2\omega)\widehat{z}$, and

$$E_{v,x}(2\omega) = \frac{k_{0,v}^2}{4\pi\epsilon_0 R}\left(p_{v,x}^{(2)}(\cos^2\theta_R + \sin^2\theta_R \sin^2\phi_R) - p_{v,y}^{(2)}\sin^2\theta_R \sin\phi_R \cos\phi_R \right.$$
$$\left. - p_{v,z}^{(2)}\sin\theta_R \cos\theta_R \cos\phi_R\right)$$

$$E_{v,y}(2\omega) = \frac{k_{0,v}^2}{4\pi\epsilon_0 R}\left(-p_{v,x}^{(2)}\sin^2\theta_R \sin\phi_R \cos\phi_R + p_{v,y}^{(2)}(\sin^2\theta_R \cos^2\phi_R + \cos^2\theta_R) \right.$$
$$\left. - p_{v,z}^{(2)}\sin\theta_R \cos\theta_R \sin\phi_R\right)$$

$$E_{v,z}(2\omega) = \frac{k_{0,v}^2}{4\pi\epsilon_0 R}\left(-p_{v,x}^{(2)}\sin\theta_R \cos\theta_R \cos\phi_R - p_{v,y}^{(2)}\sin\theta_R \cos\theta_R \sin\phi_R + p_{v,z}^{(2)}\sin^2\theta_R\right)$$

$$(7)$$

The exponential term $e^{i k_0 \cdot (R-r) + i 2 k_1 \cdot r}$ reflects the phase change of $E_v(2\omega)$ and can be rewritten as $e^{i k_0 \cdot R - i q \cdot r}$, in which $q \equiv k_0 - 2 k_1$ is the scattering wavevector.

As light is a transverse wave, at detection, the polarization state of the scattered SH field is typically analyzed in two orthogonal directions that are perpendicular to the wavevector. As illustrated in Fig. 1, we specify the directions of the analyzed polarizations as

$$\widehat{e}_1 = -(\sin\varphi_e \cos\theta_R \cos\phi_R + \cos\varphi_e \sin\phi_R)\widehat{x} + (\cos\varphi_e \cos\phi_R - \sin\varphi_e \cos\theta_R \sin\phi_R)\widehat{y}$$
$$+ \sin\varphi_e \sin\theta_R \, \widehat{z}$$

$$\widehat{e}_2 = (\cos\varphi_e \cos\theta_R \cos\phi_R - \sin\varphi_e \sin\phi_R)\widehat{x} + (\cos\varphi_e \sin\phi_R \cos\theta_R + \sin\varphi_e \cos\phi_R)\widehat{y}$$
$$- \cos\varphi_e \sin\theta_R \, \widehat{z}$$

$$(8)$$

$\varphi_e$ is the angle between the unit vector $\widehat{e}_2$ and the plane defined by the $z$-axis and $R$ as illustrated in Fig. 1. Accordingly, the scattered SH field can be expressed as the sum of the two orthogonal components along $\widehat{e}_1$ and $\widehat{e}_2$ respectively by combining Eqs. 7 and 8:

$$E_v(2\omega) = \frac{k_{0,v}^2}{4\pi\epsilon_0 R}\Big\{\Big[p_{v,x}^{(2)}(-\sin\phi_R \cos\varphi_e - \cos\theta_R \cos\phi_R \sin\varphi_e) + p_{v,y}^{(2)}(\cos\phi_R \cos\varphi_e - \cos\theta_R \sin\phi_R \sin\varphi_e) + p_{v,z}^{(2)}\sin\theta_R \sin\varphi_e\Big]\widehat{e}_1 + \Big[p_{v,x}^{(2)}(-\sin\phi_R \sin\varphi_e + \cos\theta_R \cos\phi_R \cos\varphi_e) + p_{v,y}^{(2)}(\cos\phi_R \sin\varphi_e + \cos\theta_R \sin\phi_R \cos\varphi_e) - p_{v,z}^{(2)}\sin\theta_R \cos\varphi_e\Big]\widehat{e}_2\Big\}. \quad (9)$$

For measurements of the scattered SH field in the $xz$-scattering plane $\phi_R = 0$ or $\pi$ and $\theta_R \in [0,\pi]$. This is equivalent to $\phi_R = 0$ and $\theta_R \in [-\pi,\pi]$. Then, inserting these conditions into Eq. 9, we can express the scattered SH field in the scattering plane as

$$E_v(2\omega) = \frac{k_{0,v}^2}{4\pi\epsilon_0 R}\Big\{\Big(-p_{v,x}^{(2)}\cos\theta_R \sin\varphi_e + p_{v,y}^{(2)}\cos\varphi_e + p_{v,z}^{(2)}\sin\theta_R \sin\varphi_e\Big)\widehat{e}_1 + \Big(p_{v,x}^{(2)}\cos\theta_R \cos\varphi_e + p_{v,y}^{(2)}\sin\varphi_e - p_{v,z}^{(2)}\sin\theta_R \cos\varphi_e\Big)\widehat{e}_2\Big\}. \quad (10)$$

Furthermore, for linear P and S polarization, $\varphi_e = \pi/2$, and we have $\hat{e}_1 = -\cos\theta_R \hat{x} + \sin\theta_R \hat{z}$ and $\hat{e}_2 = \hat{y}$ indicating the directions of the P and S polarizations, respectively. In this case, Eq. 9 is further simplified as

$$\boldsymbol{E}_v(2\omega) = \frac{k_{0,v}^2}{4\pi\epsilon_0 R}\left\{\left(-p_{v,x}^{(2)}\cos\theta_R + p_{v,z}^{(2)}\sin\theta_R\right)\hat{e}_1 + \left(p_{v,y}^{(2)}\right)\hat{e}_2\right\}. \quad (11)$$

Treating all molecules as individual scatterers, we can sum up all the SH fields from every molecule and then obtain the total SH intensity (which is the normally measured quantity in experiments) as[5]

$$I(2\omega) = \frac{cn\epsilon_0}{2}\left|\sum_v \widetilde{\boldsymbol{E}}_v(2\omega)\right|^2 = \frac{cn\epsilon_0}{2}\left\{\underbrace{\sum_v |\boldsymbol{E}_v(2\omega)|^2}_{\text{self-correlations, incoherent}} + \underbrace{\sum_{v\neq v'} \boldsymbol{E}_v(2\omega)\boldsymbol{E}_{v'}^*(2\omega)e^{i\boldsymbol{q}\cdot(\boldsymbol{r}-\boldsymbol{r'})}}_{\text{cross-correlations, coherent}}\right\} \quad (12)$$

where $c$ is the velocity of light in vacuum, $n$ is the refractive index of air, and $\boldsymbol{E}_{v'}^*(2\omega)$ is the complex conjugate of $\boldsymbol{E}_{v'}(2\omega)$. The first term in Eq. 12 represents self-correlations of individual molecules and is an incoherent HRS contribution to the total SH intensity; the second term, i.e. the double summation over $v$ and $v'$, represents cross-correlations between molecules and leads to a coherent contribution to the total SH intensity. The correlation between two molecules will thus obtain a phase factor $e^{i\boldsymbol{q}\cdot(\boldsymbol{r}-\boldsymbol{r'})}$, in which $\boldsymbol{q}\cdot(\boldsymbol{r}-\boldsymbol{r'})$ is the phase difference between the SH field of the two molecules. Combining Eqs. 4, 9, and 12, we can express the experimentally measured SHS intensity as a function of the molecular hyperpolarizability $\boldsymbol{\beta}^{(2)}$ of individual molecules and their correlations.

**Molecular hyperpolarizability**

$\boldsymbol{\beta}^{(2)}$ characterizes the SH response of a molecule and is intrinsically written in the molecular frame. The values of $\boldsymbol{\beta}^{(2)}$ tensor elements are determined by the nature of the second-order optical process (resonant or non-resonant) and the electronic structure of the molecule.[10] For non-resonant SHG in lossless media in which the chromatic dispersion is negligible, the values of $\boldsymbol{\beta}^{(2)}$ elements are real and the following permutation symmetry holds[1]

$$\beta_{ijk}^{(2)} = \beta_{ikj}^{(2)} = \beta_{jik}^{(2)} = \beta_{jki}^{(2)} = \beta_{kji}^{(2)} = \beta_{kij}^{(2)} \quad (13)$$

The number of independent $\boldsymbol{\beta}^{(2)}$ tensor elements can be further reduced according to the spatial symmetry of the molecule. The water molecule ($H_2O$) belongs to the point group $C_{2v}$ with two planes of symmetry, the $ac$- and $bc$-plane, and a 2-fold axis of symmetry, the $c$-axis. The coordinate system $(a,b,c)$ defines the molecular frame. As a molecular property, $\boldsymbol{\beta}^{(2)}$ of water possesses the same symmetry. Accordingly, the non-zero $\boldsymbol{\beta}^{(2)}$ elements of water are as follows and related as Eq. 13 for elastic non-resonant SHS:

$$\beta_{aac}^{(2)} = \beta_{aca}^{(2)} = \beta_{caa}^{(2)},$$

$$\beta_{bbc}^{(2)} = \beta_{bcb}^{(2)} = \beta_{cbb}^{(2)},$$

$$\beta_{ccc}^{(2)}. \quad (14)$$

It is worth noting that theoretical studies commonly assume $\boldsymbol{\beta}^{(2)}$ to be constant, independent of the environment and molecular geometry.[15-19] However, as recently reported in Ref. [7], the values of the $\boldsymbol{\beta}^{(2)}$ elements of water fluctuate significantly and show broad distributions due to the inhomogeneity of the local environment and nuclear quantum effects.

In addition, the molecular symmetry is found to be broken by fluctuations in the liquid phase. For liquid water, the instantaneous $\boldsymbol{\beta}^{(2)}$ tensor has 7 more independent non-zero elements with broad distributions around zero (thus having mean values of zero). It is therefore likely that for femtosecond experiments, distributions of the $\boldsymbol{\beta}^{(2)}$ elements rather than constant values need to be considered. We will however neglect this here.

A conversion of the above $\boldsymbol{\beta}_v^{(2)}$ elements from the molecular frame to the lab frame is needed for the calculation in Eq. 4. This conversion of $\boldsymbol{\beta}_v^{(2)}$ from the molecular frame of individual molecules to the lab frame is determined by the orientation of the molecule. As shown in Fig. 1, in the lab frame, the orientation of molecule $v$ is described by the three Euler angles $\phi_v$, $\theta_v$, and $\psi_v$. The conversion from the molecular frame $(a, b, c)$ to the lab frame $(x, y, z)$ can be given by

$$\hat{x} = (\cos\phi_v \cos\psi_v \cos\theta_v - \sin\psi_v \sin\phi_v)\hat{a} + (-\sin\phi_v \cos\psi_v - \cos\phi_v \cos\theta_v \sin\psi_v)\hat{b} + \sin\theta_v \cos\phi_v \hat{c}$$

$$\hat{y} = (\cos\psi_v \sin\phi_v \cos\theta_v + \cos\phi_v \sin\psi_v)\hat{a} + (\cos\psi_v \cos\phi_v - \sin\phi_v \sin\psi_v \cos\theta_v)\hat{b} + \sin\phi_v \sin\theta_v \hat{c}$$

$$\hat{z} = -\cos\psi_v \sin\theta_v \hat{a} + \sin\psi_v \sin\theta_v \hat{b} + \cos\theta_v \hat{c} \tag{15}$$

The 'y convention' described on page 607 in Ref. [20] is adopted here for the Euler angles.

**Calculation of the incoherent SHS intensity**

Given the large number of molecules involved in this optical process, the summation in Eq. 12 is equivalent to a statistical average over all the molecules multiplied by the number of molecules. Thus, the incoherent SHS, or equivalently HRS, intensity can be given by

$$I(2\omega) = \frac{cn\epsilon_0}{2} N_m V \langle |E_v(2\omega)|^2 \rangle \tag{16}$$

where $N_m$ is the number density of the molecule, and $V$ is the volume in which molecules contribute to the SHS intensity and typically corresponds to the focal volume of the incident laser beam. For the SHS field that is analyzed in the polarization direction of $\hat{e}_1$, the HRS intensity can be given by inserting Eq. 9 into Eq. 16:

$$I(2\omega) = \frac{cnk_0^4 N_m V}{32\pi^2 \epsilon_0 R^2} \langle \left| p_x^{(2)}(-\sin\phi_R \cos\varphi_e - \cos\theta_R \cos\phi_R \sin\varphi_e) \right.$$
$$\left. + p_y^{(2)}(\cos\phi_R \cos\varphi_e - \cos\theta_R \sin\phi_R \sin\varphi_e) + p_z^{(2)} \sin\theta_R \sin\varphi_e \right|^2 \rangle$$

$$= \frac{cnk_0^4 N_m V}{32\pi^2 \epsilon_0 R^2} \langle \left|p_x^{(2)}\right|^2 (\sin\phi_R \cos\varphi_e + \cos\theta_R \cos\phi_R \sin\varphi_e)^2 + \left|p_y^{(2)}\right|^2 (\cos\phi_R \cos\varphi_e - \cos\theta_R \sin\phi_R \sin\varphi_e)^2 + \left|p_z^{(2)}\right|^2 \sin^2\theta_R \sin^2\varphi_e -$$
$$\left[p_x^{(2)}\left(p_y^{(2)}\right)^* + p_y^{(2)}\left(p_x^{(2)}\right)^*\right](\sin\phi_R \cos\varphi_e + \cos\theta_R \cos\phi_R \sin\varphi_e)(\cos\phi_R \cos\varphi_e - \cos\theta_R \sin\phi_R \sin\varphi_e) -$$
$$\left[p_x^{(2)}\left(p_z^{(2)}\right)^* + p_z^{(2)}\left(p_x^{(2)}\right)^*\right](\sin\phi_R \cos\varphi_e + \cos\theta_R \cos\phi_R \sin\varphi_e)\sin\theta_R \sin\varphi_e +$$
$$\left[p_y^{(2)}\left(p_z^{(2)}\right)^* + p_z^{(2)}\left(p_y^{(2)}\right)^*\right](\cos\phi_R \cos\varphi_e - \cos\theta_R \sin\phi_R \sin\varphi_e)\sin\theta_R \sin\varphi_e \rangle$$

$$\tag{17}$$

where $\left(p_y^{(2)}\right)^*$ represents the complex conjugate of $\left(p_y^{(2)}\right)$. As shown in Eq. 17, to obtain the incoherent SHS intensity, it is essential to calculate ensemble average products of the Cartesian components of $\boldsymbol{p}^{(2)}$. Given the expression of $\boldsymbol{p}^{(2)}$ in Eq. 4, we can translate these products into a set of products of $\boldsymbol{\beta}^{(2)}$ elements. As expected for isotropic liquids, the molecules orient randomly with an even probability in all directions. The product $\langle\beta_{IJK}\beta_{I'J'K'}\rangle$, where the subscripts can be $x, y,$ or $z$, possesses spatial centrosymmetry. Accordingly, even numbers of the indices $x, y$, and $z$ are required for nonzero product $\langle\beta_{IJK}\beta_{I'J'K'}\rangle$. We can list the products of $\boldsymbol{p}^{(2)}$ components as follows

$$\langle|p_x^{(2)}|^2\rangle = \langle(\beta_{xyy}^{(2)})^2\rangle E_y^4 + \langle(\beta_{xxx}^{(2)})^2\rangle E_x^4 + \langle(\beta_{xyx}^{(2)} + \beta_{xxy}^{(2)})^2\rangle E_x^2 E_y^2 + 2\langle\beta_{xyy}^{(2)}\beta_{xxx}^{(2)}\rangle E_x^2 E_y^2 \cos 2\phi_0$$

$$\langle|p_y^{(2)}|^2\rangle = \langle(\beta_{yyy}^{(2)})^2\rangle E_y^4 + \langle(\beta_{yxx}^{(2)})^2\rangle E_x^4 + \langle(\beta_{yyx}^{(2)} + \beta_{yxy}^{(2)})^2\rangle E_x^2 E_y^2 + 2\langle\beta_{yyy}^{(2)}\beta_{yxx}^{(2)}\rangle E_x^2 E_y^2 \cos 2\phi_0$$

$$\langle|p_z^{(2)}|^2\rangle = \langle(\beta_{zyy}^{(2)})^2\rangle E_y^4 + \langle(\beta_{zxx}^{(2)})^2\rangle E_x^4 + \langle(\beta_{zyx}^{(2)} + \beta_{zxy}^{(2)})^2\rangle E_x^2 E_y^2 + 2\langle\beta_{zyy}^{(2)}\beta_{zxx}^{(2)}\rangle E_x^2 E_y^2 \cos 2\phi_0$$

$$\langle p_x^{(2)}(p_y^{(2)})^* + p_y^{(2)}(p_x^{(2)})^*\rangle = 2\langle\beta_{xyy}^{(2)}\beta_{yyy}^{(2)} + \beta_{xyy}^{(2)}\beta_{yxy}^{(2)} + \beta_{yyy}^{(2)}\beta_{xyx}^{(2)} + \beta_{yyy}^{(2)}\beta_{xxy}^{(2)}\rangle E_x E_y^3 \cos\phi_0 +$$
$$2\langle\beta_{xxx}^{(2)}\beta_{yyx}^{(2)} + \beta_{xxx}^{(2)}\beta_{yxy}^{(2)} + \beta_{yxx}^{(2)}\beta_{xyx}^{(2)} + \beta_{yxx}^{(2)}\beta_{xxy}^{(2)}\rangle E_x^3 E_y \cos\phi_0$$

$$\langle p_x^{(2)}(p_z^{(2)})^* + p_z^{(2)}(p_x^{(2)})^*\rangle = \langle p_z^{(2)}(p_y^{(2)})^* + p_y^{(2)}(p_z^{(2)})^*\rangle = 0 \tag{18}$$

For an isotropic medium, the lab frame indices $x, y$ and $z$ for the subscripts of $\langle\beta_{IJK}\beta_{I'J'K'}\rangle$ can be mutually permuted. There are 5 permutation ways in total:

$$\begin{cases} x \to y \\ y \to z \\ z \to x \end{cases}, \begin{cases} x \to y \\ y \to x \\ z \to z \end{cases}, \begin{cases} x \to z \\ y \to x \\ z \to y \end{cases}, \begin{cases} x \to z \\ y \to y \\ z \to x \end{cases}, \begin{cases} x \to x \\ y \to z \\ z \to y \end{cases} \tag{19}$$

Therefore, there are 10 independent $\langle\beta_{IJK}\beta_{I'J'K'}\rangle$ terms remaining in Eq. 18:

|  | $h$ | $b_1$ | $b_2$ | $b_3$ | $b_4$ | $b_5$ | $c_1$ | $c_2$ | $c_3$ | $d_1$ | $d_2$ |
|---|---|---|---|---|---|---|---|---|---|---|---|
| $\langle(\beta_{yyy}^{(2)})^2\rangle$ | $\frac{1}{7}$ | $\frac{2}{35}$ | $\frac{2}{35}$ | $\frac{1}{35}$ | $\frac{2}{35}$ | $\frac{1}{35}$ | $\frac{1}{105}$ | $\frac{1}{105}$ | $\frac{2}{105}$ | $\frac{1}{210}$ | $\frac{1}{105}$ |
| $\langle(\beta_{yxx}^{(2)})^2\rangle$ | $\frac{1}{35}$ | $\frac{4}{105}$ | $\frac{-1}{35}$ | $\frac{3}{35}$ | $\frac{-1}{35}$ | $\frac{2}{105}$ | $\frac{1}{35}$ | $\frac{-1}{210}$ | $\frac{-1}{105}$ | $\frac{1}{70}$ | $\frac{-1}{210}$ |
| $\langle\beta_{zyy}^{(2)}\beta_{zxx}^{(2)}\rangle$ | $\frac{1}{105}$ | $\frac{2}{35}$ | $\frac{-1}{105}$ | $\frac{1}{35}$ | $\frac{-1}{105}$ | $\frac{-1}{210}$ | $\frac{8}{105}$ | $\frac{1}{105}$ | $\frac{-1}{21}$ | $\frac{-1}{84}$ | $\frac{1}{105}$ |
| $\langle\beta_{yyy}^{(2)}\beta_{yxx}^{(2)}\rangle$ | $\frac{1}{35}$ | $\frac{11}{105}$ | $\frac{1}{210}$ | $\frac{2}{105}$ | $\frac{1}{210}$ | $\frac{-1}{70}$ | $\frac{1}{35}$ | $\frac{-1}{210}$ | $\frac{1}{42}$ | $\frac{-1}{420}$ | $\frac{-1}{210}$ |
| $\langle(\beta_{yyx}^{(2)} + \beta_{yxy}^{(2)})^2\rangle$ | $\frac{4}{35}$ | $\frac{-4}{35}$ | $\frac{2}{105}$ | $\frac{8}{105}$ | $\frac{2}{105}$ | $\frac{8}{105}$ | $\frac{-2}{105}$ | $\frac{1}{70}$ | $\frac{-4}{105}$ | $\frac{1}{42}$ | $\frac{1}{70}$ |
| $\langle(\beta_{zyx}^{(2)} + \beta_{zxy}^{(2)})^2\rangle$ | $\frac{4}{105}$ | $\frac{-4}{105}$ | $\frac{-4}{105}$ | $\frac{4}{35}$ | $\frac{-4}{105}$ | $\frac{1}{21}$ | $\frac{-2}{21}$ | $\frac{-1}{35}$ | $\frac{8}{105}$ | $\frac{11}{210}$ | $\frac{-1}{35}$ |

**Table 1.** Six independent $\langle\beta_{IJK}\beta_{I'J'K'}\rangle$ terms adapted from Ref. [5].

|  | $h$ | $b_1$ | $b_{21}$ | $b_{22}$ | $b_3$ | $b_{41}$ | $b_{42}$ | $b_{51}$ | $b_{52}$ | $b_{53}$ | $c_1$ | $c_{21}$ | $c_{22}$ | $c_{23}$ |
|---|---|---|---|---|---|---|---|---|---|---|---|---|---|---|
| $\langle \beta^{(2)}_{xyy} \beta^{(2)}_{yyx} \rangle$ | $\frac{1}{35}$ | $\frac{1}{210}$ | $\frac{-1}{35}$ | $\frac{1}{210}$ | $\frac{-1}{70}$ | $\frac{11}{105}$ | $\frac{1}{210}$ | $\frac{2}{105}$ | $\frac{-1}{70}$ | $\frac{1}{210}$ | $\frac{-1}{210}$ | $\frac{-1}{210}$ | $\frac{-1}{105}$ | $\frac{-1}{210}$ |
| $\langle \beta^{(2)}_{xyy} \beta^{(2)}_{yxy} \rangle$ | $\frac{1}{35}$ | $\frac{1}{210}$ | $\frac{1}{210}$ | $\frac{-1}{35}$ | $\frac{-1}{70}$ | $\frac{1}{210}$ | $\frac{11}{105}$ | $\frac{-1}{70}$ | $\frac{2}{105}$ | $\frac{1}{210}$ | $\frac{-1}{210}$ | $\frac{-1}{210}$ | $\frac{-1}{105}$ | $\frac{-1}{210}$ |
| $\langle \beta^{(2)}_{yyy} \beta^{(2)}_{xyx} \rangle$ | $\frac{1}{35}$ | $\frac{1}{210}$ | $\frac{11}{105}$ | $\frac{1}{210}$ | $\frac{-1}{70}$ | $\frac{-1}{35}$ | $\frac{1}{210}$ | $\frac{2}{105}$ | $\frac{-1}{70}$ | $\frac{1}{210}$ | $\frac{-1}{210}$ | $\frac{1}{35}$ | $\frac{1}{42}$ | $\frac{-1}{210}$ |
| $\langle \beta^{(2)}_{yyy} \beta^{(2)}_{xxy} \rangle$ | $\frac{1}{35}$ | $\frac{1}{210}$ | $\frac{1}{210}$ | $\frac{11}{105}$ | $\frac{-1}{70}$ | $\frac{1}{210}$ | $\frac{-1}{35}$ | $\frac{-1}{70}$ | $\frac{2}{105}$ | $\frac{1}{210}$ | $\frac{-1}{210}$ | $\frac{-1}{210}$ | $\frac{1}{42}$ | $\frac{1}{35}$ |

|  | $c_{31}$ | $c_{32}$ | $d_{11}$ | $d_{12}$ | $d_{21}$ | $d_{22}$ | $d_{23}$ | $d_{24}$ |
|---|---|---|---|---|---|---|---|---|
| $\langle \beta^{(2)}_{xyy} \beta^{(2)}_{yyx} \rangle$ | $\frac{-1}{105}$ | $\frac{1}{42}$ | $\frac{-1}{210}$ | $\frac{-1}{210}$ | $\frac{-1}{210}$ | $\frac{-1}{210}$ | $\frac{1}{35}$ | $\frac{1}{35}$ |
| $\langle \beta^{(2)}_{xyy} \beta^{(2)}_{yxy} \rangle$ | $\frac{1}{42}$ | $\frac{-1}{105}$ | $\frac{-1}{210}$ | $\frac{-1}{210}$ | $\frac{1}{35}$ | $\frac{1}{35}$ | $\frac{-1}{210}$ | $\frac{-1}{210}$ |
| $\langle \beta^{(2)}_{yyy} \beta^{(2)}_{xyx} \rangle$ | $\frac{1}{42}$ | $\frac{-1}{105}$ | $\frac{-1}{210}$ | $\frac{-1}{210}$ | $\frac{-1}{210}$ | $\frac{-1}{210}$ | $\frac{-1}{210}$ | $\frac{-1}{210}$ |
| $\langle \beta^{(2)}_{yyy} \beta^{(2)}_{xxy} \rangle$ | $\frac{-1}{105}$ | $\frac{1}{42}$ | $\frac{-1}{210}$ | $\frac{-1}{210}$ | $\frac{-1}{210}$ | $\frac{-1}{210}$ | $\frac{-1}{210}$ | $\frac{-1}{210}$ |

**Table 2**. The other four independent $\langle \beta_{IJK} \beta_{I'J'K'} \rangle$ terms calculated for isotropic media.

The parameters $h, b, c,$ and $d$ are defined as follows

$$h = \sum_i \left(\beta^{(2)}_{iii}\right)^2, \qquad b_1 = \sum_{i,j} \beta^{(2)}_{iii} \beta^{(2)}_{ijj}, \qquad b_2 = \sum_{i,j} \beta^{(2)}_{iii} \left(\beta^{(2)}_{jij} + \beta^{(2)}_{jji}\right),$$

$$b_{21} = \sum_{i,j} \beta^{(2)}_{iii} \beta^{(2)}_{jij}, \qquad b_{22} = \sum_{i,j} \beta^{(2)}_{iii} \beta^{(2)}_{jji}, \qquad b_3 = \sum_{i,j} \left(\beta^{(2)}_{ijj}\right)^2, \qquad b_4 = \sum_{i,j} \beta^{(2)}_{ijj} \left(\beta^{(2)}_{jij} + \beta^{(2)}_{jji}\right),$$

$$b_{41} = \sum_{i,j} \beta^{(2)}_{ijj} \beta^{(2)}_{jji}, \qquad b_{42} = \sum_{i,j} \beta^{(2)}_{ijj} \beta^{(2)}_{jij}, \qquad b_5 = \sum_{i,j} \left(\beta^{(2)}_{jij} + \beta^{(2)}_{jji}\right)^2, \qquad b_{51} = \sum_{i,j} \left(\beta^{(2)}_{jij}\right)^2,$$

$$b_{52} = \sum_{i,j} \left(\beta^{(2)}_{jji}\right)^2, \qquad b_{53} = \sum_{i,j} \beta^{(2)}_{jij} \beta^{(2)}_{jji}, \qquad c_1 = \sum_{i,j,k} \beta^{(2)}_{ijj} \beta^{(2)}_{ikk},$$

$$c_2 = \sum_{i,j,k} \left(\beta^{(2)}_{jij} + \beta^{(2)}_{jji}\right)\left(\beta^{(2)}_{kik} + \beta^{(2)}_{kki}\right), \qquad c_{21} = \sum_{i,j,k} \beta^{(2)}_{jij} \beta^{(2)}_{kik}, \qquad c_{22} = \sum_{i,j,k} \beta^{(2)}_{jji} \beta^{(2)}_{kik},$$

$$c_{23} = \sum_{i,j,k} \beta^{(2)}_{jji} \beta^{(2)}_{kki}, \qquad c_3 = \sum_{i,j,k} \beta^{(2)}_{ijj} \left(\beta^{(2)}_{kik} + \beta^{(2)}_{kki}\right), \qquad c_{31} = \sum_{i,j,k} \beta^{(2)}_{ijj} \beta^{(2)}_{kik},$$

$$c_{32} = \sum_{i,j,k} \beta^{(2)}_{ijj} \beta^{(2)}_{kki}, \qquad d_1 = \sum_{i,j,k} \left(\beta^{(2)}_{ijk} + \beta^{(2)}_{ikj}\right)^2, \qquad d_{11} = \sum_{i,j,k} \left(\beta^{(2)}_{ijk}\right)^2,$$

$$d_{12} = \sum_{i,j,k} \beta^{(2)}_{ijk}\beta^{(2)}_{ikj}, \qquad d_2 = \sum_{i,j,k}\left(\beta^{(2)}_{ijk}+\beta^{(2)}_{ikj}\right)\left(\beta^{(2)}_{jik}+\beta^{(2)}_{jki}\right), \qquad d_{21} = \sum_{i,j,k}\beta^{(2)}_{ijk}\beta^{(2)}_{jik},$$

$$d_{22} = \sum_{i,j,k}\beta^{(2)}_{ikj}\beta^{(2)}_{jik}, \qquad d_{23} = \sum_{i,j,k}\beta^{(2)}_{ijk}\beta^{(2)}_{jki}, \qquad d_{24} = \sum_{i,j,k}\beta^{(2)}_{ikj}\beta^{(2)}_{jki}.$$

(20)

The subscripts $i$ ($j,k$) refer to $a,b$, or $c$ of the molecular coordinate system. The conversion of $\boldsymbol{\beta}^{(2)}$ tensor elements from the molecular frame to the lab frame follows the rules given by Eq. 15. The coefficients listed in Table 1 and 2 are calculated by doing orientational averaging of $\langle \beta_{ijk}\beta_{i'j'k'}\rangle$ for randomly oriented molecules.

For SHS in a lossless and dispersionless medium, the relations among $\boldsymbol{\beta}^{(2)}$ tensor elements as described by Eq. 13 apply. Accordingly, the number of independent parameters $h, b, c$, and $d$ as given in Eq. 20 is dramatically reduced to 5 as follows:

$$h, \qquad b_1 = \frac{1}{2}b_2 = b_{21} = b_{22}, \qquad b_3 = \frac{1}{2}b_4 = b_{41} = b_{42} = \frac{1}{4}b_5 = b_{51} = b_{52} = b_{53},$$

$$c_1 = \frac{1}{4}c_2 = c_{21} = c_{22} = c_{23} = \frac{1}{2}c_3 = c_{31} = c_{32},$$

$$\frac{1}{4}d_1 = d_{11} = d_{12} = \frac{1}{4}d_2 = d_{21} = d_{22} = d_{23} = d_{24}.$$

(21)

And the independent $\langle \beta_{IJK}\beta_{I'J'K'}\rangle$ terms remaining in Eq. 18 are reduced to

$$\langle(\beta^{(2)}_{yyy})^2\rangle, \qquad \langle(\beta^{(2)}_{yxx})^2\rangle = \frac{1}{4}\langle(\beta^{(2)}_{yyx}+\beta^{(2)}_{yxy})^2\rangle = \langle\beta^{(2)}_{xyy}\beta^{(2)}_{yyx}\rangle = \langle\beta^{(2)}_{xyy}\beta^{(2)}_{yxy}\rangle, \qquad \langle\beta^{(2)}_{zyy}\beta^{(2)}_{zxx}\rangle,$$

$$\langle\beta^{(2)}_{yyy}\beta^{(2)}_{yxx}\rangle = \langle\beta^{(2)}_{yyy}\beta^{(2)}_{xyx}\rangle = \langle\beta^{(2)}_{yyy}\beta^{(2)}_{xxy}\rangle, \qquad \frac{1}{4}\langle(\beta^{(2)}_{zyx}+\beta^{(2)}_{zxy})^2\rangle = \langle(\beta^{(2)}_{xyz})^2\rangle.$$

(22)

Under these conditions, we thus obtain a general expression of $I(2\omega)$, in the polarization direction of $\hat{e}_1$, in terms of $\boldsymbol{\beta}^{(2)}$ tensor elements for HRS as follows

$$I(2\omega) = \frac{cnk_0^4 N_m V}{32\pi^2 \epsilon_0 R^2} \Big\{ \langle \big(\beta_{yyy}^{(2)}\big)^2 \rangle \big( E_x^4 (\sin\phi_R \cos\varphi_e + \cos\theta_R \cos\phi_R \sin\varphi_e)^2$$
$$+ E_y^4 (\cos\phi_R \cos\varphi_e - \cos\theta_R \sin\phi_R \sin\varphi_e)^2 \big)$$
$$+ \langle \big(\beta_{yxx}^{(2)}\big)^2 \rangle \big( E_x^4 (\sin\phi_R \cos\varphi_e + \cos\theta_R \cos\phi_R \sin\varphi_e)^2 + E_x^4 \sin^2\theta_R \sin^2\varphi_e$$
$$+ E_y^4 (\cos\phi_R \cos\varphi_e - \cos\theta_R \sin\phi_R \sin\varphi_e)^2 + E_y^4 \sin^2\theta_R \sin^2\varphi_e$$
$$+ 4 E_x^2 E_y^2 (\cos^2\varphi_e + \cos^2\theta_R \sin^2\varphi_e)$$
$$+ 4 E_x E_y (E_x^2 + E_y^2) \cos\phi_0 (\sin\phi_R \cos\varphi_e + \cos\theta_R \cos\phi_R \sin\varphi_e)(\cos\phi_R \cos\varphi_e$$
$$- \cos\theta_R \sin\phi_R \sin\varphi_e) \big)$$
$$+ \langle \beta_{yyy}^{(2)} \beta_{yxx}^{(2)} \rangle \big( 2 E_x^2 E_y^2 \cos 2\phi_0 (\cos^2\varphi_e + \cos^2\theta_R \sin^2\varphi_e)$$
$$+ 4 E_x E_y (E_x^2 + E_y^2) \cos\phi_0 (\sin\phi_R \cos\varphi_e + \cos\theta_R \cos\phi_R \sin\varphi_e)(\cos\phi_R \cos\varphi_e$$
$$- \cos\theta_R \sin\phi_R \sin\varphi_e) \big) + 4 \langle \big(\beta_{xyz}^{(2)}\big)^2 \rangle E_x^2 E_y^2 \sin^2\theta_R \sin^2\varphi_e$$
$$+ 2 \langle \beta_{zyy}^{(2)} \beta_{zxx}^{(2)} \rangle E_x^2 E_y^2 \cos 2\phi_0 \sin^2\theta_R \sin^2\varphi_e \Big\}$$

(23)

For the measurements of the scattered SH field in the scattering plane, i.e. the $xz$-plane, the above seemingly cumbersome expression can be largely simplified with the condition that $\phi_R = 0$ and $\theta_R \in [-\pi, \pi]$:

$$I(2\omega) = \frac{cnk_0^4 N_m V}{32\pi^2 \epsilon_0 R^2} \Big\{ \langle \big(\beta_{yyy}^{(2)}\big)^2 \rangle \big( E_x^4 (\cos\theta_R \sin\varphi_e)^2 + E_y^4 (\cos\varphi_e)^2 \big)$$
$$+ \langle \big(\beta_{yxx}^{(2)}\big)^2 \rangle \big( E_x^4 \sin^2\varphi_e + E_y^4 (\cos^2\varphi_e + \sin^2\theta_R \sin^2\varphi_e)$$
$$+ 4 E_x^2 E_y^2 (\cos^2\varphi_e + \cos^2\theta_R \sin^2\varphi_e)$$
$$+ 4 E_x E_y (E_x^2 + E_y^2) \cos\phi_0 \cos\theta_R \sin\varphi_e \cos\varphi_e \big)$$
$$+ \langle \beta_{yyy}^{(2)} \beta_{yxx}^{(2)} \rangle \big( 2 E_x^2 E_y^2 \cos 2\phi_0 (\cos^2\varphi_e + \cos^2\theta_R \sin^2\varphi_e)$$
$$+ 4 E_x E_y (E_x^2 + E_y^2) \cos\phi_0 \cos\theta_R \sin\varphi_e \cos\varphi_e \big) + 4 \langle \big(\beta_{xyz}^{(2)}\big)^2 \rangle E_x^2 E_y^2 \sin^2\theta_R \sin^2\varphi_e$$
$$+ 2 \langle \beta_{zyy}^{(2)} \beta_{zxx}^{(2)} \rangle E_x^2 E_y^2 \cos 2\phi_0 \sin^2\theta_R \sin^2\varphi_e \Big\}$$

(24)

For the routinely performed SHS measurements in the SSS, PPP, SPP, and PSS polarization combinations, the measured SHS intensity can be given by

$$I_{SSS}(2\omega) = \frac{cnk_0^4 N_m V E_y^4}{32\pi^2 \epsilon_0 R^2} \langle \big(\beta_{yyy}^{(2)}\big)^2 \rangle$$

$$I_{PPP}(2\omega) = \frac{cnk_0^4 N_m V E_x^4}{32\pi^2 \epsilon_0 R^2} \Big\{ \langle \big(\beta_{yyy}^{(2)}\big)^2 \rangle \cos^2\theta_R + \langle \big(\beta_{yxx}^{(2)}\big)^2 \rangle \sin^2\theta_R \Big\}$$

$$I_{SPP}(2\omega) = \frac{cnk_0^4 N_m V E_x^4}{32\pi^2 \epsilon_0 R^2} \langle \big(\beta_{yxx}^{(2)}\big)^2 \rangle$$

$$I_{PSS}(2\omega) = \frac{cnk_0^4 N_m V E_y^4}{32\pi^2 \epsilon_0 R^2} \langle \big(\beta_{yxx}^{(2)}\big)^2 \rangle \qquad (25)$$

The subscripts of $I(2\omega)$, from left to right, denote the polarizations of the involved beams from the high frequency to the low frequency, respectively.

**Differences with previously published work**

In doing the derivation we would find the following differences with previous work, which we list here below for completeness.

Bersohn *et al.* have several errors in their pioneering paper[5]: the expression of the scattering electric field vector (Eq. 20) is incorrect; not all the cross product of molecular dipoles equals zero; more non-zero cross product of $\boldsymbol{\beta}^{(2)}$ elements should be considered in the calculation of the HRS intensity and included in table I in Ref. [5].

In the theoretical study by Shelton[13], the discussion of the SH response of molecules ($\boldsymbol{\beta}^{(2)}$) stopped in the lab frame; there is no discussion about the link between the lab frame and individual molecules. The expressions of the two transverse-optical (TO) modes (Eqs. 7 and 8) are incorrect in Ref. [13]. The given expressions in the paper fail to satisfy the requirement of mutual orthogonality between the two TO modes and the longitudinal-optical mode (expressed in Eq. 9).

In the work of Bersohn *et al.*[5], Kauranen and Persoons[11], and Shelton[13], only the detection polarizations of the SH light in the directions that is parallel to the $xz$-scattering plane and the corresponding orthogonal direction were considered. Besides, the pre-factors in the expression of the SH intensity, which are crucial for quantitative evaluations, were either omitted or not considered in a consistent way.

For isotropic liquids and other isotropic media such as gases, we can calculate the molecular hyperpolarizability by inserting the SH intensity measured under different conditions into the above equations. On the other hand, knowing values of the $\boldsymbol{\beta}^{(2)}$ tensor elements, from such as those calculated for water molecules by ab initio simulations, we can calculate SHS patterns in different polarization combinations. Any significant deviations between the calculation and measurement data indicate the presence of coherent SHS and thus correlations between the noncentrosymmetric molecules.

**Conclusions**

In summary, we present complete expressions for HRS from isotropic liquids with more general considerations of the scattering geometry and polarization states of involved light compared to the previous theoretical studies of HRS. The table of independent $\langle \beta_{IJK}\beta_{I'J'K'} \rangle$ terms is completed by expanding from 6 terms (given in Ref. [5]) to 10 terms. The general expressions of the SH intensity (Eq. 23) are given as a function of the input optical fields, $\langle \beta_{IJK}\beta_{I'J'K'} \rangle$ of the liquid molecules, and the scattering geometry. We can use these expressions and their simplified versions (Eqs. 24 and 25) to quantitatively evaluate HRS from isotropic liquids and make comparisons with the measured quantities under various experimental conditions.

**Acknowledgement**

This work was supported by the Julia Jacobi Foundation, the Swiss National Science Foundation (grant number 200021_140472), and the European Research Council (grant number 616305). The authors thank David M. Wilkins for critical reading of the manuscript.

**References**

(1) Boyd, R. W. *Nonlinear optics*; Academic Press: New York, 2008.
(2) Roke, S.; Gonella, G. *Annu. Rev. Phys. Chem.* **2012**, *63*, 353.
(3) Eisenthal, K. B. *Chem. Rev.* **2006**, *106*, 1462.
(4) Chen, Y.; Okur, H. I.; Gomopoulos, N.; Macias-Romero, C.; Cremer, P. S.; Petersen, P. B.; Tocci, G.; Wilkins, D. M.; Liang, C.; Ceriotti, M.; Roke, S. *Science Advances* **2016**, *2*, e1501891.
(5) Bersohn, R.; Pao, Y. H.; Frisch, H. L. *J. Chem. Phys.* **1966**, *45*, 3184.
(6) Tocci, G.; Liang, C.; Wilkins, D. M.; Roke, S.; Ceriotti, M. *The Journal of Physical Chemistry Letters* **2016**, *7*, 4311.
(7) Liang, C.; Tocci, G.; Wilkins, D. M.; Grisafi, A.; Roke, S.; Ceriotti, M. *submitted* **2017**.
(8) Shelton, D. P. *Journal of Chemical Physics* **2002**, *117*, 9374.
(9) Terhune, R. W.; Maker, P. D.; Savage, C. M. *Phys. Rev. Lett.* **1965**, *14*, 681.
(10) Cyvin, S. J.; Rauch, J. E.; Decius, J. C. *J. Chem. Phys.* **1965**, *43*, 4083.
(11) Kauranen, M.; Persoons, A. *J. Chem. Phys.* **1996**, *104*, 3445.
(12) McClain, W. M. *J. Chem. Phys.* **1972**, *57*, 2264.
(13) Shelton, D. P. *J. Opt. Soc. Am. B* **2000**, *17*, 2032.
(14) Jackson, J. D. *Classical electrodynamics*; Wiley: New York, 1975.
(15) Gubskaya, A. V.; Kusalik, P. G. *Mol. Phys.* **2001**, *99*, 1107.
(16) Kongsted, J.; Osted, A.; Mikkelsen, K. V.; Christiansen, O. *The Journal of Chemical Physics* **2003**, *119*, 10519.
(17) Jensen, L.; van Duijnen, P. T.; Snijders, J. G. *The Journal of Chemical Physics* **2003**, *119*, 12998.
(18) Sylvester-Hvid, K. O.; Mikkelsen, K. V.; Norman, P.; Jonsson, D.; Ågren, H. *The Journal of Physical Chemistry A* **2004**, *108*, 8961.
(19) Shiratori, K.; Yamaguchi, S.; Tahara, T.; Morita, A. *The Journal of Chemical Physics* **2013**, *138*, 064704.
(20) Goldstein, H.; Poole, C. P.; Safko, J. L. *Classical mechanics*; 3rd ed.; Addison Wesley: San Francisco, 2002.